\documentclass[twocolumn]{svjour3} 
\def\makeheadbox{\relax}
\usepackage{pbox}

\usepackage[numbers]{natbib}
\usepackage{multirow}
\date{}
\usepackage{flushend}
\usepackage{underscore} 
\usepackage{booktabs} 
\usepackage[utf8]{inputenc}
\usepackage[ruled]{algorithm2e} 

\SetAlFnt{\small}
\SetAlCapFnt{\small}
\SetAlCapNameFnt{\small}
\SetAlCapHSkip{0pt}
\IncMargin{-\parindent}
\usepackage{grffile}
\usepackage{xargs}                     
\usepackage{xcolor}  
\usepackage[color=red,textsize=tiny,textwidth=1.4cm]{todonotes}

\usepackage{hyperref}
\usepackage{subfloat,subfig}
\usepackage{siunitx}

\usepackage{color}
\definecolor{lightgray}{rgb}{.9,.9,.9}
\definecolor{darkgray}{rgb}{.4,.4,.4}
\definecolor{purple}{rgb}{0.65, 0.12, 0.82}
\usepackage{arydshln}

\usepackage{etoolbox}
\usepackage{xcolor}
\usepackage{listings}

\newtoggle{InString}{}
\togglefalse{InString}

\lstdefinelanguage{JavaScript}{
  keywords={typeof, new, true, false, catch, function, return, null, catch, switch, var, if, in, while, do, else, case, break},
  keywordstyle=\color{blue}\bfseries,
  ndkeywords={class, export, boolean, throw, implements, import, this},
  ndkeywordstyle=\color{darkgray}\bfseries,
  identifierstyle=\color{black},
  sensitive=false,
  numberstyle=\color{red}\ttfamily,
  comment=[l]{//},
  morecomment=[s]{/*}{*/},
  commentstyle=\color{purple}\ttfamily,
  stringstyle=\color{blue}\ttfamily,
  morestring=[b]',
  morestring=[b]"
}

\usepackage{tikz}
\usetikzlibrary{trees}
\lstdefinestyle{customjavascript}{%
   language=JavaScript,
   backgroundcolor=\color{white},
   extendedchars=true,
   basicstyle=\linespread{0.4}\footnotesize\ttfamily,
   showstringspaces=false,
   showspaces=false,
   numberstyle=\footnotesize,
   tabsize=2,
   breaklines=true,
   showtabs=false}

\lstdefinestyle{customc}{%
  belowcaptionskip=1\baselineskip,
  breaklines=true,
  xleftmargin=\parindent,
  language=C,
  showstringspaces=false,
  basicstyle=\small\ttfamily,
  keywordstyle=\bfseries\color{green!40!black},
  numberstyle=\tiny,
  commentstyle=\itshape\color{purple!40!black},
  identifierstyle=\bfseries\color{black},
  stringstyle=\color{orange},
   morekeywords={uint64_t,uint32_t,__m256i,__m128i,UINT64_C},
}

\begin{document}
\title{Parsing Gigabytes of JSON per Second}

\author{Geoff Langdale        \and
        Daniel Lemire
}


\institute{Geoff Langdale \at
              branchfree.org \\
              Sydney, NSW \\
              Australia\\
              \email{geoff.langdale@gmail.com}          
           \and
           Daniel Lemire \at
              Universit\'e du Qu\'ebec (TELUQ)\\
              Montreal, Quebec \\
              Canada \\
              \email{lemire@gmail.com}
}


\maketitle
\begin{abstract}
JavaScript Object Notation or JSON is a ubiquitous data exchange format on the Web.  Ingesting JSON documents can become a performance bottleneck due to the sheer volume of data. We are thus motivated to make JSON parsing as fast as possible.

Despite the maturity of the problem of JSON parsing, we show that substantial speedups are possible.
We present the first  standard-compliant JSON parser to process gigabytes of data per second on a single core, using  commodity processors. We can use a quarter or fewer instructions than a state-of-the-art reference parser like RapidJSON\@. 
Unlike other validating parsers, our software (simdjson) makes extensive use of Single Instruction, Multiple Data (SIMD) instructions.
To ensure reproducibility, simdjson is freely available as open-source  software under a liberal license.
\end{abstract}





\section{Introduction}

JavaScript Object Notation (JSON) is a text format used to represent data~\cite{rfc8259}. It is commonly used for browser-server communication on the Web. It is supported by many database systems such as MySQL, PostgreSQL, IBM~DB2, SQL Server, Oracle, and  data-science frameworks such as Pandas. Many document-oriented databases are centered around JSON such as CouchDB or RethinkDB.   

The JSON syntax can be viewed as a restricted form of JavaScript, but it is used in many programming languages. JSON has four primitive types or atoms (string, number, Boolean, null) that can be embedded within composed types (arrays and objects). An object takes the form of a series of key-value pairs between braces, where keys are strings  (e.g., \texttt{\{"name":"Jack","age":22\}}). An array is a list of comma-separated values between brackets (e.g., \texttt{[1,"abc",null]}). Composed types can contain primitive types or arbitrarily deeply nested composed types as values. See Fig.~\ref{fig:jsonexample} for an example. The JSON specification defines six  \emph{structural characters} (`\texttt{[}', `\texttt{\{}', `\texttt{]}', `\texttt{\}}', `\texttt{:}', `\texttt{,}'): they serve to delimit the locations and structure of objects and arrays.

\begin{figure*}\centering
\begin{tabular}{c|c}
\begin{minipage}[b]{0.5\columnwidth}
   \lstset{escapechar=@,style=customjavascript}
\begin{lstlisting}
{
	"Width": 800,
	"Height": 600,
	"Title": "View from my room",
	"Url": "http://ex.com/img.png",
	"Private": false,
	"Thumbnail": {
		"Url": "http://ex.com/th.png",
		"Height": 125,
		"Width": 100
	},
	"array": [
		116,
		943,
		234
	],
	"Owner": null
}
\end{lstlisting}
\end{minipage}
&
\tikzstyle{every node}=[draw=black,thick,anchor=west]
\tikzstyle{selected}=[draw=red,fill=red!30]
\tikzstyle{optional}=[dashed,fill=gray!50]
\begin{tikzpicture}[%
  scale=0.6, transform shape,
  grow via three points={one child at (0.5,-0.7) and
  two children at (0.5,-0.7) and (0.5,-1.4)},
  edge from parent path={(\tikzparentnode.south) |- (\tikzchildnode.west)}]
  \node {root}
    child { node {"Width": 800}}		
    child { node {"Height": 600}}
    child { node {"Title": "View from my room"}}
    child { node {"Url": "http://ex.com/img.png"}}
    child { node {"Private": false}}
    child { node {"Thumbnail"}
      child { node {"Url": "http://ex.com/th.png"}}
      child { node {"Height": 125}}
      child { node {"Width": 100}}
    }
    child [missing] {}				
    child [missing] {}				
    child [missing] {}		
    child { node {"array"}
      child { node {116}}
      child { node {943}}
      child { node {234}}
    }
    child [missing] {}				
    child [missing] {}				
    child [missing] {}	
    child { node {"Owner": null}};
\end{tikzpicture}
\end{tabular}
\caption{\label{fig:jsonexample} JSON example}
\end{figure*}

To access the data contained in a JSON document from software, it is typical
to transform the JSON text into a tree-like logical representation, akin to the right-hand-side of Fig.~\ref{fig:jsonexample}, an operation we call JSON parsing.  We refer to each value, object and array as a \emph{node} in the parsed tree. After parsing, the programmer can access each node in turn and navigate to its siblings or its children without  need for complicated and error-prone string parsing.

Parsing large JSON documents is a common task. Palkar et al.\ state that big-data applications  can spend 80--90\% of their  time parsing JSON documents~\cite{palkar2018filter}.
Boncz et al.\ identified the acceleration
of JSON parsing as a topic of interest  for speeding up database processing~\cite{boncz2019database}. 

JSON parsing implies error checking: arrays must start and end with a bracket, objects must start and end with a brace, objects must be made of comma-separated pairs of values (separated by a colon) where all keys are strings. Numbers must follow the specification and fit within a valid range. Outside of string values, only a few ASCII characters are allowed. Within string values, several characters (like ASCII line endings) must be escaped. The JSON specification requires that documents use a unicode character encoding (UTF-8, UTF-16, or UTF-32), with UTF-8 being the default. Thus we must validate the character encoding of all strings. JSON parsing is therefore more onerous than merely locating nodes. Our contention is that a parser that accepts erroneous JSON is both dangerous---in that it will silently accept malformed JSON whether this has been generated accidentally or maliciously---and poorly specified---it is difficult to anticipate or widely agree on what the semantics of malformed JSON files should be.

To accelerate processing, we should
use our processors as efficiently as possible.
Commodity processors (Intel, AMD, ARM, POWER) support
single-instruction-multiple-data (SIMD) instructions. These SIMD instructions operate on several words at once unlike regular instructions. 
For example, starting with the Haswell microarchitecture (2013), Intel and AMD processors support the AVX2 instruction set and 256-bit vector registers. Hence, on recent x64 processors, we can compare two strings of 32~characters in a single instruction. 
It is thus straightforward to use SIMD instructions to locate significant characters (e.g., `\texttt{"}', `\texttt{=}') using few instructions. We refer to the application of SIMD instructions as \emph{vectorization}.
Vectorized software tends to use fewer instructions than conventional software.
Everything else being equal, code that generates fewer instructions is faster.

A closely related concept to vectorization is branchless processing: whenever the processor must choose between two code paths (a branch), there is a risk of incurring several cycles of penalty due to a mispredicted branch on current pipelined processors. In our experience, SIMD instructions are most likely to be beneficial in a branchless setting.

To our knowledge, publicly available JSON validating parsers make little use of SIMD instructions. 
Due to its complexity, the full JSON parsing problem may not appear immediately amenable to vectorization.

One of our core results is that  SIMD instructions combined with minimal branching can lead to  new speed records for JSON parsing---often processing gigabytes of data per second on a single core. We present several specific performance-oriented strategies that are of general interest.
\begin{itemize}
\item We detect quoted strings, using solely arithmetic and logical operations and a fixed number of instructions per input bytes, while omitting escaped quotes (\S~\ref{sec:escapedquotes}).  
\item We differentiate between sets of  code-point values  using vectorized classification thus avoiding the burden of doing $N$~comparisons to recognize that a value is part of a set of size $N$ (\S~\ref{sec:vectorizedclassification}).
\item We validate UTF-8 strings using solely SIMD instructions (\S~\ref{sec:charactervalidation}).
\end{itemize}

\section{Related Work}

A common strategy to accelerate JSON parsing in the literature is to parse selectively. Alagiannis et al.~\cite{Alagiannis:2012:NAA:2367502.2367543} presented NoDB, an approach where one queries the JSON data without first loading it in the database. It relies in part on selective parsing of the input.
 Bonetta and Brantner use speculative just-in-time (JIT) compilation and  selective data access to speed up JSON processing~\cite{Bonetta:2017:FFJ:3137765.3137782}. They find repeated constant structures and generate code targeting these structures. 
 
 Li et al.\ present their fast parser, Mison which can jump directly to a queried field without parsing intermediate content~\cite{Li:2017:MFJ:3115404.3115416}. Mison uses SIMD instructions to quickly identify some structural characters but otherwise works by processing bit-vectors in general purpose registers with branch-heavy loops. Mison does not attempt to validate documents; it assumes that documents are pure ASCII as opposed to unicode (UTF-8).
 We summarize the most relevant component of the Mison architecture as follows:
 \begin{enumerate}
     \item In a first step, the input document is compared against each of the structural characters (`\texttt{[}', `\texttt{\{}', `\texttt{]}', `\texttt{\}}', `\texttt{:}', `\texttt{,}') as well as the backslash (\texttt{`\textbackslash{}'}). Each comparison uses a SIMD instruction, comparing 32~pairs of bytes at a time. The comparisons are then converted into  bitmaps where the bit value~1 indicate the presence of the corresponding structural character.  Mison omits the structural characters related to arrays (`\texttt{[}', `\texttt{]}') when they are unneeded. Mison only uses SIMD instructions during this first step; it also appears to be the only step that is essentially branch-free.
    \item During a second step, Mison identifies the starting and ending point of each string in the document. It uses the quote and backslash bitmaps. For each quote character, Mison counts the number of preceding backslashes using a fast instruction (\texttt{popcnt}): quotes preceded by an odd number of backslashes are turned off and ignored.
    \item During a third step, Mison identifies the string spans delimited by the quotes. It takes each word (e.g., 32~bits) from the bitmap produced during the second step. It  iteratively turns pairs of quotes into a string mask (where a 1-bit indicates the content of a string); using a small number of arithmetic and logical operations during each iteration.
    \item In a final step, Mison uses the string masks to turn off and ignore structural characters (e.g., \texttt{\{}, \texttt{\}}, \texttt{:}) contained inside strings. Mison stores all opening braces in a stack. It pops the stack with each new closing brace, starting from the left, thus finding pairs of matching braces. For each possible nesting depth, a bitmap indicating the location of the colons can be constructed by partially copying the input colon bitmap. 
 \end{enumerate}
 Starting from the colon locations extracted from the bitmaps, Mison can parse the keys by scanning backward and the values by scanning forward. It can select only the content at a given depth. In effect, the colon bitmap serves as an index to selectively parse the input document. In some instances, Mison can scan through JSON documents at a speed of over 2\,GB/s for high selectivity queries on a 3.5\,GHz Intel processor. It is faster than what is possible with a conventional validating parser like RapidJSON\@.

FishStore~\cite{Xie:2019:FFI:3299869.3319896} parses  JSON data and selects subsets of interest, storing the result in a fast key-value store~\cite{Chandramouli:2018:FCK:3183713.3196898}. While the original FishStore relied on Mison, the open-source version\footnote{\url{https://github.com/microsoft/FishStore}} uses simdjson by default for  fast parsing.


 Pavlopoulou et al.~\cite{pavlopoulou2018parallel} propose a parallelized JSON processor that supports advanced queries and rewrite rules. It avoids the need to first load the data.
 
 Sparser filters quickly an unprocessed document to find mostly just the relevant information~\cite{palkar2018filter}, and then relies on a  parser.  We could use simdjson with Sparser.
 
 Systems based on selective parsing like Mison or Sparser might be beneficial when only a small subset of the data is of interest. However, if the data is accessed repeatedly, it might be preferable to load the data in a database engine using a standard parser. Non-validating parsers like Mison might be best with tightly integrated systems where invalid inputs are unlikely.



\subsection{XML Parsing}

Before JSON, there has been a lot of similar work done on parsing XML\@. Noga et al.~\cite{Noga:2002:LXP:585058.585075} report that when fewer than 80\% of the values need to be parsed, it is more economical to parse just the needed values. Marian et al.~\cite{Marian:2003:PXD:1315451.1315471} propose to ``project'' XML documents, down to a smaller document before executing queries.  Green et al.~\cite{Green:2004:PXS:1042046.1042051} show that we can parse XML quickly using a Deterministic Finite Automaton (DFA) where the states are computed lazily, during parsing. Farf\'an et al.~\cite{Farfan:2007:BLX:2395856.2395868} go further and skip entire sections of the XML document, using  internal physical pointers. Takase et al.~\cite{Takase:2005:AFS:1060745.1060845} accelerate XML parsing by avoiding syntactic analysis when subsets of text have been previously encountered. Kostoulas et al.\ designed a fast validating XML parser called Screamer: it achieves higher speed by reducing the number of distinct processing steps~\cite{Kostoulas:2006:XSI:1135777.1135796}. 
Cameron et al.\ show that we can parse XML faster using SIMD instructions~\cite{Cameron:2008:HPX:1463788.1463811}, in their parser (called Parabix). 
Zhang et al.~\cite{zhang2009speculative} show how we can parse XML documents in parallel by first indexing the document, and then separately parsing partitions of the document.

Mytkowicz et al.~\cite{newMytkowicz:2014:DFM:2541940.2541988} show how to vectorize finite-state machines using SIMD instructions. They demonstrate good results with HTML tokenization, being more than twice as fast as a baseline.

\subsection{CSV Parsing}

Data also comes in the form of comma-separated values (CSV). M\"uhlbauer et al.\ optimize CSV parsing and loading using SIMD instructions to locate delimiters and invalid characters~\cite{Muhlbauer:2013:ILM:2556549.2556555}.
Ge et al.\ use a two-pass approach where the first pass identifies the regions between delimiters while the second pass processes the records~\cite{speculative-distributed-csv-data-parsing-for-big-data-analytics}.





\section{Parser Architecture and Implementation}


In our experience, most JSON parsers proceed by top-down recursive descent~\cite{Cohen:1978:ADP:359511.359517} that makes a single pass through the input bytes, doing character-by-character decoding. We adopt a different strategy, using two distinct passes. 
We briefly describe the two stages before covering them in detail in subsequent sections.

\begin{enumerate}
\item 
In stage 1, we validate the character encoding and identify  the starting location of all JSON nodes (e.g., numbers, strings, null, true, false, arrays, objects).
We also need the location of all structural characters (`\texttt{[}', `\texttt{\{}', `\texttt{]}', `\texttt{\}}', `\texttt{:}', `\texttt{,}') defined in the JSON specification~\cite{rfc8259}. 
These locations are written as integer indexes in a separate array. 

During this stage, it is necessary to distinguish the characters that are between quotes, and thus inside a
string value, from other characters. For example,
the JSON document \texttt{"[1,2]"} is a single string despite the appearance of brackets. That is, these brackets should not be identified as relevant structural characters.
Because quotes
can be escaped (e.g.,  `\texttt{\textbackslash{}"}'),
it is necessary to identify backslash characters as well. Outside of strings, only four specific white-space characters are allowed (space, tab, line feed, carriage return). Any other white-space character needs to be identified.

The first stage involves either SIMD processing over bytes or the manipulation of bitsets (arrays of bits) that have one~bit corresponding to one~byte of input. As such, it can be  inefficient for some inputs---we can observe dozens of operations taking place to discover that there are in fact no odd-numbered sequences of backslashes or quotes in a given block of input. However, this inefficiency on such inputs is balanced by the fact that it costs no more to run this code over complex structured input, and the alternatives would generally involve running a number of unpredictable branches. 

\item In stage~2, we process all of the nodes and structural characters. We distinguish the nodes based on their starting character. When a quote (`\texttt{"}') is encountered, we parse a string; when a digit or a hyphen is found, we parse a number; when the letters `\texttt{t}', `\texttt{f}', `\texttt{n}' are found, we look for the values \texttt{true}, \texttt{false} and \texttt{null}. 

Strings in JSON cannot contain 
some characters unescaped, i.e., ASCII characters with code points less than 0x20, and they may contain many sorts of escaped characters. It is thus necessary to \emph{normalize} the strings: convert them to valid UTF-8 sequences.

Encountered numbers must be converted to either  integers or  floating-point values. They can take many forms (e.g., \texttt{12}, \texttt{3.1416}, \texttt{1.2e+1}). However, we must check many rules while parsing numbers. For example, the following strings are invalid numbers: \texttt{012}, \texttt{1E+}, and \texttt{.1}. The JSON specification is not specific regarding the range of numbers that we should accept: many parsers cannot 
represent integer values outside of the interval $[-2^{53},2^{53}]$. In contrast, we choose
to accept all 64-bit integers in the interval $[-2^{63}, 2^{63})$, while rejecting integers outside this interval. We also reject overly large floating-point numbers (e.g, \texttt{1e309} or \texttt{-1e309}).

We validate objects as sequences of strings, colons (`\texttt{:}') and values; we validate arrays as sequences of values separated by commas (`\texttt{,}'). We ensure that all objects started with an open brace (`\texttt{\{}') are terminated with a closing brace (`\texttt{\}}').  We ensure that all arrays started with an open square bracket (`\texttt{[}') are terminated with a closing square bracket ( `\texttt{]}').

The result is written  in document order on  a \emph{tape}: an array of 64-bit words. The tape contains a word for each node value (string, number, true, false, null) and a word at the beginning and at the end of each object or array.  To ensure fast navigation, the words on the  tape corresponding to braces or brackets are annotated so that we can go from the word at the start of an object or array to the word at the end of the array without reading the content of the array or object. 
Specifically, the tape is constructed as follows.
\begin{itemize}
    \item A \texttt{null} atom is represented as the 64-bit value (\texttt{'n'}$\times 2^{56}$) where `n' is the 8-bit code point values (in ASCII) corresponding to the letter `n'. A \texttt{true} atom is represented by \texttt{'t'}$\times 2^{56}$, a  \texttt{false} atom is given by \texttt{'f'}$\times 2^{56}$.
\item Numbers are represented using two 64-bit words. Integers are given by the word \texttt{'l'}$\times 2^{56}$ followed by a 64-bit signed integer (in standard two's complement form). Floating-point numbers are given by the word \texttt{'d'}$\times 2^{56}$ followed by a 64-bit floating point value (in standard binary64).
\item For an array, the first 64-bit tape element contains the value \texttt{'['}$\times 2^{56}+x$ where $x$ is one plus the index of the second 64-bit tape element on the tape.
The second 64-bit tape element contains the value \texttt{']'}$\times 2^{56}+x$ where $x$ is the index of the first 64-bit tape element on the tape. 
All the content of the array is located between these two tape elements.
We proceed similarly with objects using words of the form \texttt{'\{'}$\times 2^{56}+x$ and  \texttt{'\}'}$\times 2^{56}+x$. In-between these two tape elements, we alternate between a key (which must be a string) and a value. A value could be an atom, an object or an array.
\item We have a secondary array (\emph{a string buffer}) where normalized string values are stored. 
We represent a string as the word
\texttt{'"'}$\times 2^{56}+x$ where $x$ is the index of the beginning of the string on the secondary array.
\end{itemize}
Additionally, we add a special word at the beginning and the end of the document. The first word is annotated to point at the last word on the tape. See Fig.~\ref{fig:jsontape} for an example of the resulting tape.

\end{enumerate}

At the end of the two stages, we report whether the JSON document is valid~\cite{rfc8259}. We produce an error code (e.g., \textsc{string_error}, \textsc{number\_error}, \textsc{utf8\_error}, etc.). All strings are normalized and all numbers have been parsed and validated.

\begin{figure}\centering
\begin{minipage}[b]{0.8\columnwidth}
   \lstset{escapechar=@,style=customjavascript}
\begin{lstlisting}
0  : r	// pointing to 37 (right after last node)
1  : {	// pointing to next tape location 37 (first node after the scope)
2  : string "Width"
3  : integer 800
5  : string "Height"
6  : integer 600
8  : string "Title"
9  : string "View from my room"
10 : string "Url"
11 : string "http://ex.com/img.png"
12 : string "Private"
13 : false
14 : string "Thumbnail"
15 : {	// pointing to next tape location 25 (first node after the scope)
16 : string "Url"
17 : string "http://ex.com/th.png"
18 : string "Height"
19 : integer 125
21 : string "Width"
22 : integer 100
24 : }	// pointing to previous tape location 15 (start of the scope)
25 : string "array"
26 : [	// pointing to next tape location 34 (first node after the scope)
27 : integer 116
29 : integer 943
31 : integer 234
33 : ]	// pointing to previous tape location 26 (start of the scope)
34 : string "Owner"
35 : null
36 : }	// pointing to previous tape location 1 (start of the scope)
37 : r	// pointing to 0 (start root)
\end{lstlisting}
\end{minipage}
\caption{\label{fig:jsontape} JSON tape corresponding to the example in Fig.~\ref{fig:jsonexample}}
\end{figure}

Our two-stage design is motivated by performance concerns. Stage~1 operates directly on the input bytes, processing the
data in batches of 64~bytes. In this manner, we can make full use of the SIMD instructions that are key to our good performance. Stage~2 is easier because stage~1 identified the location of all atoms, objects and arrays: there is no need to finish parsing one atom to know the location of the next one.
Thus when parsing \texttt{\{"key1":"val1", "key2":"val2"\}}, stage~2 receives the location of the tokens \texttt{\}}, \texttt{"}, \texttt{:}, \texttt{"}, \texttt{"}, \texttt{:}, \texttt{"}, \texttt{\}} corresponding to the start and end of the object, the two colons, and the beginning of each of the four strings. It can process the four strings without data dependence---we do not need to complete the  parsing of the string \texttt{"key1"} to find the location of the column (\texttt{:}) before finding the location of the string  \texttt{"val1"}.
Except for unicode validation, we deliberately delay number and string validation to stage~2, as these tasks are comparatively expensive and difficult to perform unconditionally and cheaply over our entire input. 

\subsection{Stage 1: Structural and Pseudo-Structural Elements}

The first stage of our processing must identify key points in our input: the \emph{structural characters} of JSON (brace, bracket, colon and comma), the start and end of strings as delineated by double quote characters, other JSON \emph{atoms} that are not distinguishable by simple characters ( \texttt{true}, \texttt{false}, \texttt{null} and numbers), as well as discovering these characters and atoms in the presence of both quoting conventions and backslash escaping conventions.

In JSON, a first pass over the input can efficiently  discover the significant characters that delineate syntactic elements (objects and arrays). Unfortunately,  these characters may also appear between quotes, so we need to identify quotes.
It is also necessary to identify the backslash character because JSON allows escaped characters: `\textbackslash{}"', `\textbackslash{}\textbackslash{}', `\textbackslash{}/', `\textbackslash{}b', `\textbackslash{}f', `\textbackslash{}n', `\textbackslash{}r', `\textbackslash{}t', as well as escaped unicode characters (e.g. \texttt{\textbackslash{}uDD1E}).



A point of reference is Mison~\cite{Li:2017:MFJ:3115404.3115416}, a fast parser in C++. 
Mison uses vector instructions to  identify the colons, braces, quotes and backslashes. The detected quotes and backslashes are used to filter out the insignificant colons and braces. 
We follow the broad outline of the construction of a structural index as set forth in Mison; first, the discovery of odd-length sequences of backslash characters---which will cause quote characters immediately following to be escaped and not serve their quoting role but instead be literal characters, second, the discovery of quote pairs---which cause structural characters within the quote pairs to also be merely literal characters and have no function as structural characters, then finally the discovery of structural characters not contained within the quote pairs.
We depart from the Mison paper in method and overall design. The Mison authors loop over the results of their initial SIMD identification of characters, while we propose branchless sequences to accomplish similar tasks. For example, to locate escaped quote characters, they iterate over the repeated quote characters. Their Algorithm~1 identifies the location of the quoted characters by iterating through the unescaped quote characters. We have no such loops in our stage~1: it is essentially branchless, with a fixed cost per input bytes (except for character-encoding validation, \S~\ref{sec:charactervalidation}). Furthermore, Mison's vectorized processing is more limited by design as it does not identify the locations of the atoms, it does not process the white-space characters and it does not validate the character encoding.

\subsubsection{Identification of the quoted substrings} 
\label{sec:escapedquotes}
\begin{figure*}
   \centering
\lstset{escapechar=@,style=customjavascript}
\centering
\begin{minipage}[b]{0.95\textwidth}
\begin{lstlisting}
{ "\\\"Nam[{": [ 116,"\\\\" , 234, "true", false ], "t":"\\\"" }: input data
___111________________1111_______________________________111____: B
1_1_1_1_1_1_1_1_1_1_1_1_1_1_1_1_1_1_1_1_1_1_1_1_1_1_1_1_1_1_1_1_: E (constant)
_1_1_1_1_1_1_1_1_1_1_1_1_1_1_1_1_1_1_1_1_1_1_1_1_1_1_1_1_1_1_1_1: O (constant)
// identify 'starts' - backslashes characters not preceded by backslashes
___1__________________1__________________________________1______: S = B &~(B << 1)

// detect end of a odd-length sequence of backslashes starting on an even offset
// detail: ES gets all 'starts' that begin on even offsets
______________________1_________________________________________: ES = S & E
// add B to ES, yielding carries on backslash sequences with even starts
___111____________________1______________________________111____: EC = B + ES
// filter out the backslashes from the previous addition, getting carries only
__________________________1_____________________________________: ECE = EC & ~B
// select only the end of sequences  ending on an odd offset
________________________________________________________________: OD1 = ECE & ~E

// detect end of a odd-length sequence of backslashes starting on an odd offset
// details are as per the above sequence
___1_____________________________________________________1______: OS = S & O
______1_______________1111__________________________________1___: OC = B + OS
______1_____________________________________________________1___: OCE = OC & ~B
______1_____________________________________________________1___: OD2 = OCE & E

// merge results, yielding ends of all odd-length sequence of backslashes
______1_____________________________________________________1___: OD = OD1 | OD2
\end{lstlisting}
\end{minipage}
    \caption{Branchless code sequence to identify escaped quote characters (with example). We use the convention of the C language: `\texttt{\&}' denotes the bitwise AND, `\texttt{|}' the bitwise OR, `\texttt{<<}' is a left shift, `\texttt{\textasciitilde{}}' is a bitwise negation.}
    \label{fig:identifyescapedquotes}
\end{figure*}

Identifying escaped quotes is less trivial than it appears. While it is easy to recognize that
the string ``\texttt{\textbackslash{}"}'' is made of an escaped quote since a quote character  immediately
preceded by a backslash, if a quote is preceded by an even number of backslashes (e.g.,  ``\texttt{\textbackslash{}\textbackslash{}"}''),
then it is not escaped since  \texttt{\textbackslash{}\textbackslash{}} is an escaped backslash.
We distinguish sequences of backslash characters starting at an odd index location
from sequences starting at even index location. A sequence of characters that starts at an odd (resp.\ even) index location
and ends at an odd (resp.\ even) index location must have an even length, and it is therefore a sequence of escaped
backslashes. Otherwise, the sequence contains an odd number of backslashes and any quote character following it
must be considered escaped. We provide the code sequence with an example in Fig.~\ref{fig:identifyescapedquotes} where two quote characters are escaped.\footnote{We simplify this sequence for clarity. 
Our results are affected by the previous iteration over the preceding 64~byte input if any. Suppose a single backslash 
ended the previous 64~byte input; this alters the results of the previous algorithm. We similarly elide the full details 
of the adjustments for previous loop state in our presentation of subsequent algorithms.} 

With the backslash and quote characters identified, we can locate the unescaped quote characters efficiently. We compute a shift followed by a bitwise ANDNOT, eliminating the \emph{escaped} quote characters. 

However, we are interested in finding the location between quotes (the strings), so we can find the actual structural characters. The desired bit pattern would be 1 if there are an odd-numbered number of unescaped quotes at or before our location and zero otherwise.  For example, given the word 0b100010000 representing quote locations (with 1-bit), we wish to compute 0b011110000.\footnote{We use the convention that 0b100010000 is the binary value with the fifth and ninth least significant bits set to 1. }  We can achieve this result using  the prefix sum of the XOR operation over our bit vector representing unescaped quotes. That is, the resulting bit value at index $i$ is the XOR of all bit values up to and including the bit value at index $i$ in the input. We can compute such a prefix sum in C++ with a loop that repeatedly apply the bitwise XOR on  a left-shifted word:\lstinline[style=customc]!for (i=0;i<64;i++) {mask = mask xor (mask << 1)}!. This prefix sum can  be more efficiently implemented as one instruction by using the carry-less multiplication~\cite{lemire2016faster} (implemented with the \texttt{pclmulqdq} instruction) of our unescaped quote bit vector by another 64-bit word made entirely of ones. 
The carry-less multiplication works like the regular integer multiplication, but, as the name suggests, without a carry because it relies on the XOR operation instead of the addition. Let us use the convention 
that given a 64-bit integer $a$, $a_i$ is the value of the $i^{\mathrm{th}}$~bit so that $a = \sum_{i=0}^{63} a_i 2^i$. The regular product between two 64-bit integers $a, b$ is given by $\sum_{i=0}^{63} a_i  b  2^i $ where $a_i  b  2^i $ is zero when $a_i$ is zero, and otherwise it is $b$ left righted by $i$~bits. With these conventions, the carry-less product is given by 
$\bigoplus_{i=0}^{63} a_i  b  2^i $; that is, we replace the sum ($\sum$) by a series of XOR (symbolized by $\bigoplus$). Thus we see that when $a_i=1$ for all indexes $i$, we get $\bigoplus_{i=0}^{63} b  2^i $ which is the prefix sum of the XOR operation. The carry-less multiplication is broadly supported and fast on recent processors due to its applications in cryptography. On skylake processors,  the carry-less multiplication (\texttt{pclmulqdq}) has a latency of 7~cycles and one can be issued per cycle~\cite{fog2018instruction}.


\begin{figure*}
    \centering
\lstset{escapechar=@,style=customjavascript}
\begin{minipage}[b]{0.95\textwidth}
\begin{lstlisting}
{ "\\\"Nam[{": [ 116,"\\\\" , 234, "true", false ], "t":"\\\"" }: input data
__1___1_____1________1____1________1____1___________1_1_1___11__: Q
______1_____________________________________________________1___: OD 
__1_________1________1____1________1____1___________1_1_1____1__: Q &= ~OD
__1111111111_________11111_________11111____________11__11111___: CLMUL(Q,~0)
\end{lstlisting}
\end{minipage}
    \caption{Branchless code sequence to identify quoted range excluding the final quote character. CLMUL refers to the carry-less multiplication. We assume that OD is computed using the code sequence from Fig.~\ref{fig:identifyescapedquotes}.}
    \label{fig:identifyquoteregion}
\end{figure*}

\subsubsection{Vectorized Classification} 
\label{sec:vectorizedclassification}
Mison does one SIMD comparison per  character (`\texttt{:}', `\texttt{\textbackslash{}}', `\texttt{"}', `\texttt{\{}', `\texttt{\}}'). 
We proceed similarly to identify the quotes
and the backslash characters. However, there are six structural characters, and, for purposes of further analysis, we also need to discover the four permissible white-space characters. Doing
ten comparisons and accompanying bitwise OR operations would be expensive.
Instead of a comparison, we use the AVX2 \texttt{vpshufb} instruction to acts as a vectorized table lookup to do a \emph{vectorized classification}~\cite{simdbase64}. The \texttt{vpshufb} instruction uses the least significant 4~bits of each byte (low nibble) as an index into a 16-byte table. 
Other processor architectures (ARM and POWER) have similar SIMD instructions.

By doing one lookup, followed by a 4-bit right shift and a second lookup (using a different table), we can separate the characters into 
one of two categories: structural characters and white-space characters. The first lookup maps the low nibbles (least significant 4~bits) of each byte to a byte value; the second lookup maps the high nibble (most significant 4~bits) of each byte  to a byte value. The two byte values are combined with a bitwise AND. 

To see how this can be used to identify sets of characters, suppose that we want to identify the byte values 0x9, 0xa and 0xd. The  low nibbles are 9, a and d, and the high nibbles are all zeroes. In the first 16-byte lookup table, we set the fourth least significant bit to 1 for the values corresponding to indexes 9, a and d, and only for these three values. In the second 16-byte lookup table,  set the fourth least significant bit to 1 for the value at index 0, and only for this value. Then we have that whenever the input values 0x9, 0xa and 0xd are encountered, and only for these values, the fourth least significant bit of the result of the bitwise AND is 1. Hence, using two \texttt{vpshufb} instructions, a shift and a few bitwise logical operations, we can identify a set of characters. If we could only identify one set of characters, this approach would not be necessarily advantageous, but we can identify many different sets with the same two \texttt{vpshufb} instructions.
We can repeat the same strategy with new sets of input values, always making them match a given bit index (the fourth in our example). To avoid misclassifications, we  need to ensure that each set of input values corresponding to a bit index is uniquely characterized by a set of low nibbles and high nibbles. The set \{0x9, 0xa, 0xd\} works since it is the set of all values with low nibbles 9, a, d, and high nibble 0.
The set \{0x5b,0x5d, 0x7b, 0x7d\} also works since it is the set of all values with the low nibbles  b and d and the high nibbles 5 and 7. The set \{0x21, 0x33\} would not work 
since 0x31 and 0x23 also share the same high nibbles and low nibbles: we would need to break the set \{0x21, 0x33\} into two subsets (\{0x21\}, \{0x33\}). We can support as many sets as we have bit indexes (8).

\begin{table*}\centering
\begin{minipage}[!t]{0.4\textwidth}
\begin{tabular}{cc|ccccccccc}
\toprule
  & \shortstack{low\\nibble}  & 0 & $\cdots$ &  9 & a & b & c & d & e & f\\
\shortstack{high\\ nibble}  &   & 16& $\cdots$ & 8  & 10& 4 & 1 & 12& 0 & 0\\\midrule
0 & 8 &   & $\cdots$ & 8 &  8 &   &   &  8&   &  \\
1 & 0 &   & $\cdots$ &   &    &   &   &   &   &  \\
2 &17 & 16& $\cdots$ &   &    &   & 1 &   &   &  \\
3 & 2 &   & $\cdots$ &   &  2 &   &   &   &   &  \\
4 & 0 &   & $\cdots$ &   &    &   &   &   &   &  \\
5 & 4 &   & $\cdots$ &   &    & 4 &   & 4 &   &  \\ 
6 & 0 &   & $\cdots$ &   &    &   &   &   &   &  \\
7 & 4 &   & $\cdots$ &   &    & 4 &   & 4 &   &  \\ \bottomrule
\end{tabular}
\end{minipage}
\hspace{0.1\textwidth}
\begin{minipage}[!t]{0.4\textwidth}
\begin{tabular}{cc}\toprule
code points  & desired value \\ \midrule
0x2c  & 1 \\[1em]
0x3a &  2 \\[1em]
0x5b,0x5d, 0x7b, 0x7d & 4 \\[1em]
0x09, 0x0a, 0x0d  & 8 \\[1em]
0x20  & 16 \\[1em]
others &  0 \\
\bottomrule
\end{tabular}
\end{minipage}
    \caption{Table describing the vectorized classification of
    the code points.
    The first column and first row are indexes corresponding to the high and low nibbles.
    The second column and the second row are the looked up table values. The main table values are
    the bitwise AND result of the two table values (e.g., 10\texttt{ AND }8 is 8). 
    The omitted values are zeroes. On the right, we give the desired classification.}
    \label{tab:vectable}
\end{table*}

We break the set of code-point values corresponding to structural characters into three sets: \{0x2c\}, \{0x3a\}, \{0x5b,0x5d, 0x7b, 0x7d\}. We match them to the first three bit indexes.
We break the set of code-point values corresponding to white-space characters into two sets \{0x9, 0xa, 0xd\}, and \{0x20\}. We match them to the fourth and fifth bit indexes. These sets are all uniquely characterized by their low and high nibbles. Our solution is not unique: e.g., we could have broken \{0x5b,0x5d, 0x7b, 0x7d\} into two sets ( \{0x5b,0x5d\}, \{0x7b, 0x7d\}). 

See Table~\ref{tab:vectable}.
The table for the low nibbles is 16, 0, 0, 0, 0, 0, 0, 0, 0, 8, 10, 4, 1, 12, 0, 0; and  the table for the high nibbles is 8, 0, 17, 2, 0, 4, 0, 4, 0, 0, 0, 0, 0, 0, 0, 0.
Applying our algorithm, we get the following results:
\begin{itemize}
\item For the comma (code-point value 0x2c), we get 1, as the bitwise AND between 17 and 1.
\item For the colon (code-point value 0x3a), we get 0b10 (2 in binary).
\item For the other structural characters `[', `]', `\{', `\}' (code-point values 0x5b,0x5d, 0x7b, 0x7d), we get 0b100 (4 in binary).
    \item For the first three white-space characters (with code-point values 0x9, 0xa and 0xd), we get the value 0b1000 (8 in binary).
    \item For the space character  (code-point value 0x20), we get 0b10000 (16 in binary).
    \item All other character inputs will yield zero.
\end{itemize}
We can recognize the structural characters (`,', `:', `[', `]', `\{', `\}') by computing a bitwise AND with 0b111 and the white-space characters with a bitwise AND with 0b11000.
That is, with only two \texttt{vpshufb} instructions and a few logical instructions, we can classify all code-point values into one of three sets: structural (comma, colon, braces, brackets), ASCII white-space (`\textbackslash{}r', `\textbackslash{}n', `\textbackslash{}t', ` ') and others. No branching is required. 

\subsubsection{Identification of White-Space and Pseudo-Structural Characters} 

We also make use of our ability to quickly detect white space in this early stage. We can use another bitset-based transformation to discover locations in our data that follow a structural character or quote followed by zero or more characters of white space; excluding locations within strings, and the structural characters we have already discovered, these locations are the only place that we can expect to see the starts of the JSON \emph{atoms} (such as numbers whether or not starting with a minus sign, \texttt{null}, \texttt{true}, and \texttt{false}). These locations are thus treated as \emph{structural} and we term them \emph{pseudo-structural characters}.
Formally, we define pseudo-structural characters as non-white-space characters that are \begin{enumerate}
\item outside quotes and
\item have a predecessor that is either a white-space character or a structural character.
\end{enumerate}
We use a feature of JSON: the legal atoms can all be distinguished from each other by their first character: `\texttt{t}' for \texttt{true}, `\texttt{f}' for \texttt{false}, `\texttt{n}' for \texttt{null} and the character class \texttt{[0-9-]} for numerical values.

As a side-effect, identifying pseudo-structural characters helps validate documents. For example, only some ASCII white-space characters are allowed unescaped outside a quoted range in JSON\@.
An isolated disallowed character would be flagged as a pseudo-structural character and subsequently rejected in stage 2. Furthermore, dangling atoms are automatically identified (as the \texttt{a} in  \texttt{[12 a]}) and will be similarly rejected. The key insight is that stage 1 need not discover whether such out-of-place characters are legal JSON---it only needs to \emph{expose} them in the stream of structural and pseudo-structural characters that will be parsed in stage 2.

Fig.~\ref{fig:identifypseudo} illustrates a final sequence in stage~1 where given the (unescaped) quotes and the quoted ranges, as well as the structural and white-space characters, we identify the pseudo-structural characters. 

\begin{figure*}
    \centering
\lstset{escapechar=@,style=customjavascript}
\begin{minipage}[b]{0.95\textwidth}
\begin{lstlisting}
{ "\\\"Nam[{": [ 116,"\\\\" , 234, "true", false ], "t":"\\\"" }: input data
__1_________1________1____1________1____1___________1_1_1____1__: Q 
__1111111111_________11111_________11111____________11__11111___: R
1_________11_1_1____1_______1____1_______1_______11____1_______1: S
_1____________1_1__________1_1____1_______1_____1__1__________1_: W
// eliminate quoted regions from our structural characters
1____________1_1____1_______1____1_______1_______11____1_______1: S = S&~R
// restore ending quotes to our structural characters 
// (for purposes of building pseudo-structural characters)
1_1_________11_1____11____1_1____1_1____11_______11_1_111____1_1: S = S|Q
// begin to calculate pseudo-structural characters
// initially; pseudo-structural characters are structural or white space
111_________11111___11____1111___111____111_____11111_111____111: P = S|W
// now move our mask for candidate pseudo-structural characters forward by one
_111_________11111___11____1111___111____111_____11111_111____11: P = P<<1
// eliminate white-space and quoted characters from our candidates
_____________1_1_1__________1_1__________1_1_____11____1_______1: P &= ~W&~R
// merge pseudo-structural characters into structural character mask
1_1_________11_1_1__11____1_1_1__1_1____11_1_____11_1_111____1_1: S = S|P
// eliminate ending quotes from our final structural characters
1_1__________1_1_1__11______1_1__1_1_____1_1_____11_1__11______1: S&~(Q&~R)
\end{lstlisting}
\end{minipage}
    \caption{Branchless code sequence to identify the structural and pseudo-structural characters. The quotes Q and the quoted range R are computed using the code sequence from Fig.~\ref{fig:identifyquoteregion}.
    The structural (S) and white-space (W) characters are identified vectorized classification.
    }
    \label{fig:identifypseudo}
\end{figure*}

\subsubsection{Index Extraction}
\label{sec:indexextract}
During stage~1, we process blocks of 64~input bytes. The end product is a 64-bit bitset with the bits corresponding to a structural or pseudo-structural characters set to 1.
Our structural and pseudo-structural characters are relatively rare and can sometimes, but not always, be infrequent. E.g., we can construct plausible JSON inputs that have such a character once ever 40~characters or once every 4~characters. As such, continuing to process the structural characters as bitsets involves manipulating data structures that are unpredictably spaced. We choose to transform these bitsets into indexes. That is, we seek a list of the  locations of the 1-bits.
Once we are done with the extraction of the indexes, we can discard the bitset. In contrast, Mison does not have such an extraction step and iterates directly over the 1-bits.


Our  implementation involves a transformation of bitsets to indexes by use of the \emph{count trailing zeroes} operation (via the \texttt{tzcnt} instruction) and an operation to clear the lowest set bit: \texttt{s = s \& (s - 1)} in C which compiles to a single  instruction (\texttt{blsr}). This strategy introduces an unpredictable branch; unless there is a regular pattern in our bitsets, we would expect to have at least one branch miss for each word. However, we employ a technique whereby we extract 8~indexes from our bitset unconditionally, then ignore any indexes that were extracted excessively by means of overwriting those indexes with the next iteration of the index extraction loop. See Fig.~\ref{fig:extraction}. This means that as long as the frequency of our set bits is below 8~bits out of 64 we expect few unpredictable branches.
The choice of the number~8 is a heuristic based on our experience with JSON documents; a larger unconditional extraction procedure would be more expensive due to having to use more operations, but even less likely to cause a branch miss as a wider range of bit densities could be handled by extracting, say, 8~indexes from our bitset.

\begin{figure}
\centering
\begin{minipage}[b]{0.95\columnwidth}
\lstset{escapechar=@,style=customc}
\begin{lstlisting}
// we decode the set bits from 's' 
// to array 'b'
uint64_t s = ...
uint32_t * b = ...
// net line => popcnt instruction
uint32_t cnt = popcount(s); 
uint32_t next_base = b + cnt;
while (s) {
  // next line => tzcnt instruction
  *b++ = idx + trailingzeroes(s); 
  // next line => blsr instruction 
  s = s & (s - 1); 
  *b++ = idx + trailingzeroes(s);
  s = s & (s - 1);
  *b++ = idx + trailingzeroes(s);   
  s = s & (s - 1);
  *b++ = idx + trailingzeroes(s);
  s = s & (s - 1);
  *b++ = idx + trailingzeroes(s);
  s = s & (s - 1);
  *b++ = idx + trailingzeroes(s);  
  s = s & (s - 1);
  *b++ = idx + trailingzeroes(s);
  s = s & (s - 1);
  *b++ = idx + trailingzeroes(s); 
  s = s & (s - 1);
}
b = new_base;
\end{lstlisting}
\end{minipage}
\caption{\label{fig:extraction}Code sequence to extract set bits out of a bitset}
\end{figure}

\subsubsection{Character-Encoding Validation}
\label{sec:charactervalidation}
In our experience,  JSON documents are served using the unicode format UTF-8. Some programming languages like Java use UTF-16 for in-memory strings. Yet  if we consider the popular JSON Java library Jackson, then the common way to serialize an object to JSON is to call the function \texttt{ObjectMapper.writeValue}, and the result is in UTF-8.
Indeed, the JSON specification indicates that many implementations do not support encodings other than UTF-8.  
Parsers like Mison assume that the character encoding is ASCII~\cite{Li:2017:MFJ:3115404.3115416}. Though it is reasonable, a safer assumption is that unicode (UTF-8) is used. Not all sequences of bytes are valid UTF-8 and thus a validating parser needs to ensure that the character encoding is correct. We assume that the incoming data is meant to follow  UTF-8, and that the parser should produce UTF-8 strings.

UTF-8 is an ASCII superset. The ASCII characters can be represented using a single byte, as a numerical value called \emph{code point} between 0 and 127 inclusively. That is, ASCII code points are an 8-bit integer with the most significant bit set to zero. UTF-8 extends
these 128~code points to a total of 
1,114,112~code points. Non-ASCII code~points are represented using  from two to four bytes, each with the most significant bit set to one.
Non-ASCII code~points cannot contain ASCII characters: we can therefore remove from   an UTF-8 stream of bytes any number of ASCII characters without affecting its validation.

Outside of strings in JSON, all characters must be ASCII\@. Only the strings require potentially expensive validation. However, there may be many small strings in a document, so it is unclear whether vectorized unicode validation would be beneficial at the individual string level. Thus we validate the input bytes as a whole.

We first test if a block of 64~bytes is made entirely of ASCII characters. It suffices to verify
that the most significant bit of all bytes is zero.  This optimization might trigger some unpredictable branches, but given how frequently JSON documents might be almost entirely composed of ASCII characters, it is a necessary risk.

If there are non-ASCII characters, we apply a vectorized UTF-8 validation algorithm.  We need to check that sequences of bytes are made of valid UTF-8 code points (see Tables~\ref{tab:unicode}).   It involves several steps, but each one is efficient. We work exclusively with SIMD instructions. 
\begin{itemize}
\item We need to verify that all byte values are no larger than 0xF4 (or 244): we can achieve this check with an 8-bit \emph{saturated} subtraction with 0xF4. The result of the subtraction is zero if and only if the value is no larger than 0xF4.

\item  When the byte value 0xED is found, the next byte must be no larger than 0x9F; when the byte value 0xF4 is found, the next byte must be no larger than 0x8F. We can check these conditions with vectorized byte comparisons and byte shifts.
\item  The byte values 0xC0 and 0xC1 are forbidden. When the byte value is 0xE0, the next byte value is larger than 0xA0. When the byte value is 0xF0, the next byte value is at least 0x90.
\item  When a byte value is outside the range of ASCII values,
it belongs to one out of four classes, depending on the value of its high nibble:
\begin{itemize}
    \item If the high nibble is 8, 9, a or b (in hexadecimal)  then the byte is the second, third of fourth byte in a code point.
    \item If the high nibble is c or d then the byte must be the first of a sequence of two bytes forming a code point.
    \item If the high nibble is e then the byte is the first out of a code point made of three bytes.
    \item Finally, if the high nibble is f, then the byte is first in a sequence of four bytes.
\end{itemize}
We use the \texttt{vpshufb} instruction to quickly map bytes to one of these categories using values 0, 2, 3, and 4. We map ASCII characters to the value 1.
If the value 4 is found (corresponding to a nibble value of f), it should be followed by three values 0. 
Given such a vector of integers, 
we can check that it matches a valid sequence of code points in the following manner. Shift values by~1 and subtract~1 using saturated subtraction, add the result to the original vector. Repeat the same process with a factor of two: shift values by~2 and subtract~2, add the result to the original vector. Starting with the sequence 
4 0 0 0 2 0 1 1 3 0 0, you first get
4 3 0 0 2 1 1 1 3 2 0 and then 
4 3 2 1 2 1 1 1 3 2 1. If the sequence came from valid UTF-8, all final values should be greater than zero, and be no larger than the original vector.
\end{itemize}
All these checks are done using SIMD registers solely, without branching.  At the beginning of the processing, we initialize an \emph{error variable} (as a 32-byte vector) with zeroes. We compute in-place the bitwise OR of the result of each check with our error variable. Should any check fail, the error variable will become non-zero. We only check at the end of the processing (once) that the variable is zero. If a diagnosis is required to determine where the error occurs, we can do a second pass over the input.



\begin{table*}\centering
\begin{tabular}{c|llll}
\toprule
code points & 1st byte & 2nd byte & 3rd byte & 4th byte \\\midrule
  \texttt{0x000000\ldots{}0x00007F}    & 00\ldots{}7F& & & \\
  \texttt{0x000080\ldots{}0x0007FF}   &  C2\ldots{}DF &  80\ldots{}BF& & \\
  \texttt{0x000800\ldots{}0x000FFF}   &  E0   &    A0\ldots{}BF  & 80\ldots{}BF& \\
  \texttt{0x001000\ldots{}0x00CFFF}   &  E1\ldots{}EC  & 80\ldots{}BF  & 80\ldots{}BF& \\
  \texttt{0x00D000\ldots{}0x00D7FF}   &  ED       & 80\ldots{}9F  & 80\ldots{}BF& \\
  \texttt{0x00E000\ldots{}0x00FFFF}   &  EE\ldots{}EF   & 80\ldots{}BF  & 80\ldots{}BF& \\
  \texttt{0x010000\ldots{}0x03FFFF}  & F0       & 90\ldots{}BF  & 80\ldots{}BF &  80\ldots{}BF \\
  \texttt{0x040000\ldots{}0x0FFFFF}  & F1\ldots{}F3   & 80\ldots{}BF  & 80\ldots{}BF &  80\ldots{}BF \\
 \texttt{0x100000\ldots{}0x10FFFF} & F4       & 80\ldots{}8F  & 80\ldots{}BF &  80\ldots{}BF   \\  \bottomrule
\end{tabular}
    \caption{Unicode code points and their representation into sequences of up to four bytes under UTF-8.}
    \label{tab:unicode}
\end{table*}

\subsection{Stage 2: Building the Tape}

In the final stage, we iterate through the indexes found in the first stage. To handle objects and arrays that can be nested, we use a goto-based state machine. Our state is recorded as a stack indicating whether we are in an array or an object,  we append our new state to the stack whenever we encounter an embedded array or object. When the embedded object or array terminates, we use the stored state from the stack and a goto command to resume the parsing from the appropriate state in the containing scope. Values such as \texttt{true}, \texttt{false}, \texttt{null} are handled as simple string comparisons. We parse numbers and strings using dedicated functions. 
Without much effort, we could support streaming processing without materializing JSON documents  objects as  in-memory tapes~\cite{Liu:2014:JDM:2588555.2595628}.

\subsubsection{Number Parsing}

It is difficult to do number parsing and validation without proceeding 
in a standard character-by-character manner.
We must check for all of the rules of the specification~\cite{rfc8259}.
Thus we proceed as do most parsers.
However, we found it useful to test for the common case where there are at least eight digits as part of the fractional portion of the number. Given the eight characters interpreted as a 64-bit integer \texttt{val}, we can check whether it is made of eight digits with an inexpensive comparison:
\begin{lstlisting}
(((val & 0xF0F0F0F0F0F0F0F0) 
| (((val + 0x0606060606060606)  
      & 0xF0F0F0F0F0F0F0F0) >> 4)) 
      == 0x3333333333333333).
\end{lstlisting}
When this check is successful, we  invoke a fast vectorized function to compute the equivalent integer value (see Fig.~\ref{fig:intval}). This  fast function begins by substracting from all character values the code point of the character `\texttt{0}', using the 
\texttt{_mm_sub_epi8} intrinsic (using the \texttt{psubb} instruction). Because the
digits have consecutive code points in ASCII, this ensures that digit characters are
mapped to their values: `\texttt{0}' becomes 0, `\texttt{1}' becomes 1 and so forth.
We then invoke the \texttt{_mm_maddubs_epi16} intrinsic (i.e., the \texttt{pmaddubsw} instruction) to multiply every other digit by 10 and add the result to the previous digit, as a 16-bit sum.
We repeat a similar process with the \texttt{_mm_madd_epi16} intrinsic (i.e., the \texttt{pmaddwd} instruction), this time multiplying every other value by 100 and adding it to the previous value as a 32-bit sum. The maximal value of these sums is 9999 which fits in a 16-bit integer. We apply the \texttt{_mm_packus_epi32} intrinsic (i.e., the \texttt{packusdw} instruction) to pack the four 32-bit integers into four 16-bit integers.
Finally, we call the \texttt{_mm_madd_epi16} intrinsic again to multiply every other 16-bit value by 10000 and add it to the preceding value, generating a 32-bit sum. We only return one 32-bit value corresponding to 8~digits even though, technically, our function converts two series of eight digits to 32-bit integers. On a Skylake processor, the instructions have a latency of 1~cycle (\texttt{psubb}, \texttt{packusdw}) or 5~cycles (\texttt{pmaddubsw} and \texttt{pmaddwd}). Because the  \texttt{pmaddwd} instruction instruction is needed twice, we can estimate that this function has a latency of at least 17~cycles, not counting store and load instructions. However, only about seven instructions are needed (with load/store instructions) which might be less than half what a naive approach might require. Indeed, merely loading eight digits may require eight distinct load instructions.

\begin{figure*}
\centering
\begin{minipage}[b]{0.99\textwidth}
\lstset{escapechar=@,linewidth=0.8\textwidth,style=customc}
\begin{lstlisting}
uint32_t parse_eight_digits_unrolled(char *chars) {
  __m128i ascii0 = _mm_set1_epi8('0');
  __m128i mul_1_10 =
      _mm_setr_epi8(10, 1, 10, 1, 10, 1, 10, 1, 10, 1, 10, 1, 10, 1, 10, 1);
  __m128i mul_1_100 = _mm_setr_epi16(100, 1, 100, 1, 100, 1, 100, 1);
  __m128i mul_1_10000 =
      _mm_setr_epi16(10000, 1, 10000, 1, 10000, 1, 10000, 1);
  __m128i in = _mm_sub_epi8(_mm_loadu_si128((__m128i *)chars), ascii0);
  __m128i t1 = _mm_maddubs_epi16(in, mul_1_10);
  __m128i t2 = _mm_madd_epi16(t1, mul_1_100);
  __m128i t3 = _mm_packus_epi32(t2, t2);
  __m128i t4 = _mm_madd_epi16(t3, mul_1_10000);
  return _mm_cvtsi128_si32(t4); 
}
\end{lstlisting}
\end{minipage}
\caption{\label{fig:intval}Code sequence using Intel intrinsics to convert eight digits to their integer value.}
\end{figure*}




\subsubsection{String Validation and Normalization}

When encountering a quote character, we always read 32~bytes in a vector register, then look for the quote and the escape characters. If an escape character is found before the first quote character, we use a conventional code path to process the escaped character, otherwise we just write the 32-byte register to our string buffer. Some unicode characters may be represented using  one or two 4-character hexadecimal code (e.g., \textbackslash{}uD834\textbackslash{}uDD1E); we convert them to valid UTF-8 bytes.

Our string buffer is made of a 32-bit integer indicating the length of the string followed by the string content in UTF-8.  The JSON specification allows strings containing null characters, hence the need for a integer specifying the length of the string. 
As part of the string validation, we must check that no code-point value less than 0x20 is found: we use vectorized comparison.

\section{Experiments}

We validate our results through a set of reproducible experiments over varied data.\footnote{Scripts, code and raw  results are available online: \url{https://github.com/lemire/simdjson} and \url{https://github.com/lemire/simdjson_experiments_vldb2019}.} In \S~\ref{sec:wherethecycles}, we report that the running time during parsing is split evenly between our two stages. In \S~\ref{sec:fewerins}, we show that we use half as many instructions during parsing as our best competitor.
In \S~\ref{sec:speedcomp}, we show that  this reduced instruction count translates into a comparable runtime advantage.

\subsection{Hardware and Software}

\label{sec:software}
Most recent Intel processors are based on the Skylake microarchitecture. We also include a computer with the more recent Cannon~Lake microarchitecture in our tests.
We summarize the characteristics of our hardware platforms in
 Table~\ref{tab:test-cpus}. We verified the CPU characteristics experimentally with \texttt{avx-turbo}~\cite{avxturbo}.
 
\begin{table*}
\caption{\label{tab:test-cpus} Hardware 
}
\centering
\begin{minipage}{\textwidth}
\centering
\begin{tabular}{cccccc}\toprule
Processor   & Base Frequency & Max. Frequency  & Microarchitecture                           & Memory  & Compiler\\ \midrule
Intel i7-6700  & 3.4\,GHz   & 3.7\,GHz & Skylake (x64, 2015) & DDR4 (2133\,MT/s) & GCC 9.1   \\
Intel i3-8121U & 2.2\,GHz  & 3.2\,GHz & Cannon~Lake (x64, 2018) & LPDDR4 (3200\,MT/s)  & GCC  9.1 \\
\bottomrule
\end{tabular}
\end{minipage}
\end{table*}

Our software was written using C++17. We tested it under several recent compilers (LLVM's clang, GNU GCC, Microsoft Visual Studio). Our library is well tested and included in actual systems such as Yandex Clickhouse.
We use the GNU GCC~9.1 compiler with the \texttt{-O3} and \texttt{-march=native} flags  under Linux. We compile the code as is, without profile-guided optimization. All code is single-threaded. We disable hyper-threading.

Our experiments assume that the JSON document is in memory; we omit disk and network accesses. Popular disks (e.g., NVMe) have a bandwidth of 3\,GB/s~\cite{Xu:2015:PAN:2757667.2757684} and more. In practice, JSON documents are frequently ingested from the network. Yet current networking standards allow for speeds exceeding 10\,GB/s~\cite{cole2011100} and modern networking hardware can allow network data to be read directly into a cache line. A high performance implementation of JSON scanning is desirable even for data  coming  from a fast disk or from the network. While we focus on speed, we also expect that more efficient parsers reduce energy  consumption.

After reviewing several parsers, we selected RapidJSON and sajson, two open-source C++ parsers, as references (see Table~\ref{tab:test-parsers}).
Palkar et al.\ describe RapidJSON as  \emph{the fastest traditional state-machine-based parser available}~\cite{palkar2018filter}. In practice, we find that another C++ parser, sajson, is faster. They are both mature and highly optimized: they were created in 2011 and 2012 respectively. The sajson parser can be used with either static or dynamic memory allocations: the static version is faster, so we adopt it.

Counting our own parser (simdjson), all three parsers can parse 64-bit floating-point numbers as well as integers. However, sajson only supports 32-bit integers whereas both RapidJSON and simdjson support 64-bit integers. RapidJSON represents overly large integers as 64-bit floating-point numbers (in a lossy manner) whereas both our parser (simdjson) and sajson reject documents with integers that they cannot exactly represent.

RapidJSON can either normalize strings in a new buffer or within the input bytes (\emph{insitu}). We find that the parsing speed is greater in \emph{insitu} mode, so we  present these better numbers. In contrast, sajson only supports \emph{insitu} parsing. Our own parser does not modify the input bytes: it has no \emph{insitu} mode.
All three parsers do UTF-8 validation of the input. 
However, the sajson parser does partial UTF-8 validation, accepting invalid code point sequences such 
as \texttt{0xb1 0x87}.
The number parsing accuracy of all three parsers is within one unit in the last place~\cite{Goldberg:1991:CSK:103162.103163}.

We consider other open-source parsers but we find that they are either  slower than RapidJSON, or that they failed to abide by the JSON specification (see \S~\ref{sec:speedcomp}). For example, parsers like gason, jsmn and ultrajson accept \texttt{[0e+]} as valid JSON\@. Parsers like fastjson and ultrajson accept unescaped line breaks  in strings.  Other parsers are tightly integrated into larger frameworks, making it difficult to benchmark them fairly. For methodological simplicity, we also do not consider parsers written in Java or other languages.

RapidJSON has compile-time options to enable optimized code paths making use of SIMD optimizations: these optimizations skip spaces between values or structural characters. However, we found both of these compile-time macros  (RAPIDJSON\_SSE2 and RAPIDJSON\_SSE42) to be systematically detrimental to performance in our tests. Moreover, they are disabled by default in the library. Thus we do not make use of these optimizations.

Other than RapidJSON, we find that none of the libraries under consideration make deliberate use of SIMD instructions. However, we expect that all libraries benefit of SIMD instructions in our tests: many functions from the standard libraries are vectorized, and the compiler translates some conventional code to SIMD instructions (e.g., via autovectorization~\cite{naishlos2004autovectorization}).

We cannot directly compare with Mison since their software is not available publicly~\cite{Li:2017:MFJ:3115404.3115416}.
However, the authors of  Mison reports speeds up to slightly over 2\,GB/s on an Intel Xeon Broadwell-EP (E5-1620 v3) with a base frequency of 3.5\,GHz: e.g., while parsing 
partially Twitter data. Mison does not attempt to validate the documents nor to parse them entirely but it also does some additional processing to answer a specific query.

\begin{table*}
\caption{\label{tab:test-parsers} Competitive parsers 
}
\centering
\begin{minipage}{\textwidth}
\centering
\begin{tabular}{lll}\toprule
Processor   & snapshot   & link  \\ \midrule
simdjson  & June 12th 2019 & \url{https://github.com/lemire/simdjson}   \\
RapidJSON  & version 1.1.0 & \url{https://github.com/Tencent/rapidjson}   \\
sajson & September 20th 2018  & \url{https://github.com/chadaustin/sajson} \\
\bottomrule
\end{tabular}
\end{minipage}
\end{table*}

\subsection{Datasets}

Parsing speed is necessarily dependent on the content of the JSON document. For a fair assessment, we chose a wide range of documents. See Table~\ref{tab:test-datasets} for detailed statistics concerning the chosen files.
 In Table~\ref{tab:test-datasets-size}, we present the number of bytes 
of both the original document and the version without extraneous white-space characters outside strings.

From the author of RapidJSON\footnote{\url{https://github.com/miloyip/nativejson-benchmark}}, we acquired three data files. We have  canada.json which is a description of the Canadian contour in GeoJSON: it contains many numbers. We have citm\_catalog.json which is commonly used benchmark file. Finally, we have twitter.json which is the result of a search for the character one in Japanese and Chinese using the Twitter API: it contains many non-ASCII characters.
From  the author of sajson\footnote{\url{https://github.com/chadaustin/sajson/tree/master/testdata}}, we retrieved several more files:  apache\_builds.json, github\_events.json,  instruments.json,  mesh.json,  mesh.pretty.json,  update-center.json.

We also generated number.json as a large  array of random floating-point numbers. We also created twitterescaped.json  which is a minified version of the twitter.json where all non-ASCII characters have been escaped.

Many of these documents  require much number parsing or much string normalization. We deliberately did not consider small documents (smaller than 50\,kB). The task of parsing many tiny documents is outside our scope.

\begin{table*}
\caption{\label{tab:test-datasets} Datasets statistics. The second last column (struct.)  is the number of structural and pseudo-structural characters. The last column is the ratio of the total number of bytes over the number of structural and pseudo-structural characters.
}
\centering\small
\begin{tabular}{l|rrrrrrrrrrrr}\toprule
file & integer & float &  string  &  non-ascii  &  object & array & null &  true & false  & struct. &  byte/struc.\\ \midrule
apache_builds & 2 & 0 & 5289  & 0 & 884 & 3 & 0 & 2 & 1 & 12365 & 10.3 \\ 
canada & 46 & 111080 & 12 & 0  & 4 & 56045 & 0 & 0 & 0  & 334374 & 6.7 \\ 
citm_catalog & 14392 & 0 & 26604 & 348 & 10937 & 10451 & 1263 & 0 & 0 &135991 & 12.7\\ 
github_events & 149 & 0 & 1891  & 4 & 180 & 19 & 24 & 57 & 7 & 4657& 14.0\\
gsoc-2018 & 0 & 0 & 34128  & 0 & 3793 & 0 & 0 & 0 & 0 &  75842 & 43.9 \\ 
instruments & 4935 & 0 & 6889  & 0 & 1012 & 194 & 431 & 17 & 109 & 27174& 8.1 \\
marine_ik & 130225 & 114950 & 38268  & 0 & 9680 & 28377 & 0 & 6 & 0 & 643013 & 4.6\\ 
mesh & 40613 & 32400 & 11 & 0 & 3 & 3610 & 0 & 0 & 0 & 153275 & 4.7\\
mesh.pretty & 40613 & 32400 & 11  & 0 & 3 & 3610 & 0 & 0 & 0 & 153275 & 10.3\\
numbers & 0 & 10001 & 0 & 0 & 0 & 1 & 0 & 0 & 0 & 20004&7.5\\
random & 5002 & 0 & 33005& 103482 & 4001 & 1001 & 0 & 495 & 505 &  88018 & 5.8\\
twitterescaped & 2108 & 1 & 18099 & 0 & 1264 & 1050 & 1946 & 345 & 2446 &  55264 & 10.1\\
twitter & 2108 & 1 & 18099  & 95406 & 1264 & 1050 & 1946 & 345 & 2446 & 55264 & 11.4\\
update-center & 0 & 0 & 27229 & 49 & 1896 & 1937 & 0 & 134 & 252 & 63420 & 8.4\\\bottomrule
\end{tabular}
\end{table*}

\begin{table*}
\caption{\label{tab:test-datasets-size} Datasets sizes: minified size omits white-space characters outside quotes.}
\centering\small
\begin{tabular}{l|rr|r}\toprule
file & bytes (minified) & bytes (original) & ratio \\ \midrule
apache_builds & 94653 &127275  & 74\% \\
canada & 2251027 & 2251027 & 100\% \\
citm_catalog & 500299 & 1727204 & 29\% \\
github_events & 53329 & 65132 & 82\% \\
gsoc-2018 & 3073766 & 3327831 & 92\% \\ 
instruments & 108313 & 220346 & 49\% \\
marine_ik & 1834197 & 2983466 & 61\% \\ 
mesh &  650573 & 723597 & 90\% \\
mesh.pretty & 753399 &  1577353  & 48\% \\
numbers & 150121 & 150124 & 100\% \\
random &461466 & 510476 & 90\% \\
twitterescaped & 562408 & 562408  & 100\% \\
twitter & 466906 & 631514 & 74\% \\
update-center & 533177 & 533178 & 100\%\\\bottomrule
\end{tabular}
\end{table*}

\subsection{Running Time Distribution\label{sec:wherethecycles}}

\begin{figure*}\centering
\subfloat[Skylake ]{
\includegraphics[width=0.49\textwidth]{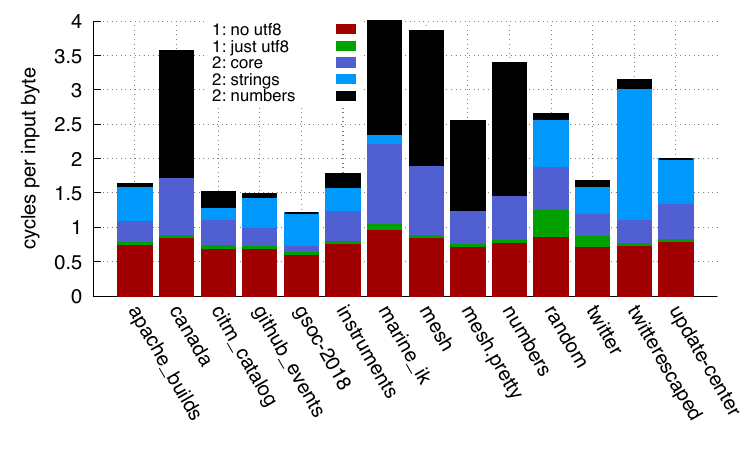}
}\subfloat[Cannon~Lake]{
\includegraphics[width=0.49\textwidth]{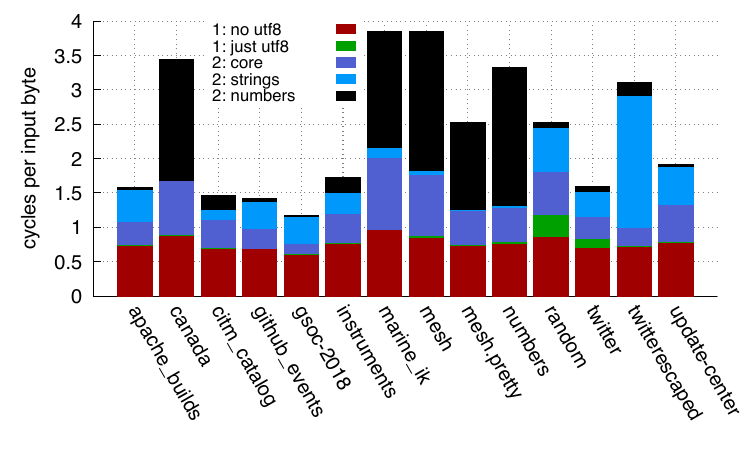}}
\caption{\label{fig:stacked}Time required in cycles per input byte to process our test files, timings are decomposed in the time needed to execute various components of the parsing. }
\end{figure*}

In Fig.~\ref{fig:stacked}, we present the distribution of cycles per stage, for each test file:
\begin{itemize}
\item The label \texttt{1: no utf8} refers to the time spent in stage~1, except for UTF-8 validation. This time is similar over all files, between 0.5 and 1~cyles per input byte. The time per input byte is higher for documents with many  structural and pseudostructural characters per input byte such as canada, marine_ik, mesh and random (see Table~\ref{tab:test-datasets}). The difference can be explained by the cost of index extraction which is higher on a per byte basis in documents with many structural characters (see \S~\ref{sec:indexextract}).
\item The label \texttt{1: just utf8} refers to the time spent doing UTF-8 validation. It is negligible for all but the random and twitter documents. 
In the random  file, UTF-8 validation is a significant cost. This file has a relatively high fraction of non-ASCII characters (20\%). In comparison, the twitter file has only 3\% of non-ASCII characters.
\item The label \texttt{2: core} is for the time spent in stage~2 except for string and number parsing. It is higher for documents marine_ik and mesh that have many structural and pseudostructural characters per input byte.
\item The label \texttt{2: strings} \texttt{2: numbers} refers to the time spent in number parsing in stage~2. Roughly a third of the CPU cycles are spent parsing numbers in the files canada, marine_jk, mesh, mesh.pretty and numbers. In other files, the time spent parsing numbers is negligible.
\item The label \texttt{2: strings} refers to the time spent parsing strings in stage~2. The string parsing time is a sizeable cost in the twitterescaped file. In this file, all non-ASCII characters have been escaped which makes string normalization more difficult.
\end{itemize}
About half the CPU cycles per input byte (between 0.5 and 3~cycles) are spent in stage~1. Much of this stage relies on  vector processing (with SIMD instructions). 
Thus  at least half of the processing time is directly related to SIMD instructions and branchless processing.

The running time of the parser depends on the characteristics of the JSON document. It can be useful to model the performance: e.g., an engineer could predict the execution time and budget accordingly. For this purpose, we used linear regression based on our dataset of files. Our dataset is relatively small, but we expect that it is large enough for a simple linear regression. 
\begin{itemize}
\item Let $F$ the number of floating-point numbers, 
\item $S$ be the number of structural and semi-structural elements and 
\item $B$ the number of bytes. 
\end{itemize}
With a high accuracy ($R^2 \geq 0.99$), we have the following cost models:
\begin{itemize}
    \item The stage~1 running time (in CPU cycles) is $1.7 \times S + 0.62 \times B$ on Skylake and 
    $1.7 \times S + 0.59 \times B$ on Cannon~Lake.
    \item The stage~2 running time is 
    $19  \times F + 8.7 \times S + 0.31 \times B$ on Skylake and 
    $17  \times F + 9.1 \times S + 0.29 \times B$
on Cannon~Lake.
\item The total running time is
$19 \times F + 11 \times S + 0.92 \times B$ on Skylake and
$17 \times F + 11 \times S + 0.88 \times B$ on Cannon~Lake.
\end{itemize}

The number of input bytes has a small coefficient (less than 1~cycle per input byte) but its contribution to the cost is still significant because there are many more bytes than structural elements or floating-point numbers.

In our model, a floating-point number not only has a direct cost, but also generates one pseudo-structural character and comprises several bytes: thus each number costs dozens of cycles.

\subsection{Fewer Instructions\label{sec:fewerins}}

The main benefit of SIMD instructions is to do more work with fewer instructions. Thus we expect our parser to user fewer instructions. We present in Table~\ref{tab:inspercycle} the number of instructions needed to parse various file using our three competitive parsers. We provide the results derived from the Skylake processors, but the number of instructions is practically the same on the Cannon~Lake processor.

On average, simdjson uses about half as many instructions as sajson and  RapidJSON\@.
Our fastest competitor, sajson, uses between  1.5 and 3.5~times more instructions. RapidJSON uses between 1.8 and 4.7~times more instructions. The files where our advantage over sajson is relatively  smallest are marine\_ik, mesh.pretty and twitterescaped. These files involve either much number parsing, or expensive string normalization.

The simdjson parser uses fewer instructions overall, but it uses many more vector instructions.
Table~\ref{tab:instructions} provides the latency and throughput of  the 256-bit vector instructions in simdjson. Most of them have minimal latency (1~cycle) and several instructions have high throughput: they can be issued twice per cycle (\texttt{vpaddb}, \texttt{vpand}, \texttt{vpcmpeqb}, etc.).
All these vector instructions can be issued at least once per cycle.

\begin{table*}
\caption{\label{tab:instructions}256-bit SIMD instructions: their latency (in cycles) and throughput (cycles/instruction) on Skylake processors~\cite{fog2018instruction}.}
\centering\small
\begin{tabular}{llrr}\toprule
Instruction & description &latency & throughput \\ \midrule
\texttt{vpaddb} &  Add  8-bit integers&    1& 0.5\\
\texttt{vpalignr} & Concatenate pairs of 16-byte blocks & 1 &     1\\
\texttt{vpand}   & Compute the bitwise AND    &  1 &    0.5\\
\texttt{vpcmpeqb}  & Compare  8-bit integers ($=$)&   1      & 0.5\\
\texttt{vpcmpgtb} & Compare  8-bit integers ($>$)    &  1    & 0.5\\
\texttt{vperm2i128} &  Shuffle integers &  3  &    1\\
\texttt{vpmaxub} & Compute max of 8-bit integers  &    1     &   0.5\\
\texttt{vpor}  & Compute the bitwise OR  &    1    &    0.5\\
\texttt{vpshufb}  & Shuffle bytes &    1     &    1\\
\texttt{vpsrld} &  Right shift  32-bit integers   &     1       &    1\\
\texttt{vpsrlw} &  Right shift  16-bit integers &     1     &        1\\
\texttt{vpsubusb}  &  Subtract  8-bit integers &    1  &   0.5\\
\texttt{vptest} &  Test for zero    &       3      &      1\\ 
\bottomrule
\end{tabular}
\end{table*}

\begin{table*}
\caption{\label{tab:inspercycle}Instructions per byte required to parse and validate documents. 
}
\centering
\begin{minipage}{\textwidth}
\centering
\begin{tabular}{lp{0.13\textwidth}p{0.13\textwidth}p{0.13\textwidth}|p{0.13\textwidth}p{0.13\textwidth}}\toprule
\multirow{2}{*}{file}   & \multirow{2}{*}{simdjson}   & \multirow{ 2}{*}{RapidJSON} & \multirow{ 2}{*}{sajson} & RapidJSON/ & sajson/  \\
   &    &  &  & simdjson & simdjson  \\
\midrule
apache_builds	&	5.6	&	15.9	&	10.0	&	2.8	&	1.8\\
canada	&	12.9	&	26.2	&	20.9	&	2.0	&	1.6\\
citm_catalog	&	5.3	&	11.7	&	11.1	&	2.2	&	2.1\\
github_events	&	4.9	&	15.5	&	10.1	&	3.2	&	2.1\\
gsoc-2018	&	3.2	&	15.0	&	11.2	&	4.7	&	3.5\\
instruments	&	6.4	&	15.3	&	12.6	&	2.4	&	2.0\\
marine_ik	&	13.4	&	23.7	&	20.6	&	1.8	&	1.5\\
mesh.pretty	&	9.0	&	17.0	&	14.9	&	1.9	&	1.7\\
mesh	&	14.3	&	27.2	&	23.3	&	1.9	&	1.6\\
numbers	&	11.7	&	25.9	&	18.8	&	2.2	&	1.6\\
random	&	8.9	&	19.6	&	15.4	&	2.2	&	1.7\\
twitter	&	5.5	&	14.3	&	11.5	&	2.6	&	2.1\\
twitterescaped	&	9.3	&	16.5	&	13.7	&	1.8	&	1.5\\
update-center	&	6.2	&	18.4	&	12.1	&	3.0	&	2.0\\
\midrule
average  &  8.3	&	18.7	&	14.7    &      &   \\\midrule
geometric mean  &      &      &         &2.4	&	1.9 \\
\bottomrule
\end{tabular}
\end{minipage}
\end{table*}

\subsection{Speed Comparison\label{sec:speedcomp}}

\begin{figure*}\centering
\subfloat[Skylake (3.4\,GHz) ]{
\includegraphics[width=0.49\textwidth]{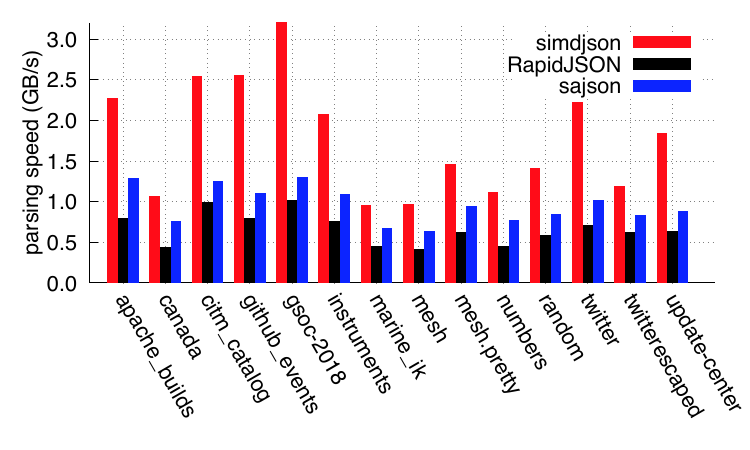}}
\subfloat[Cannon~Lake (2.2\,GHz)]{
\includegraphics[width=0.49\textwidth]{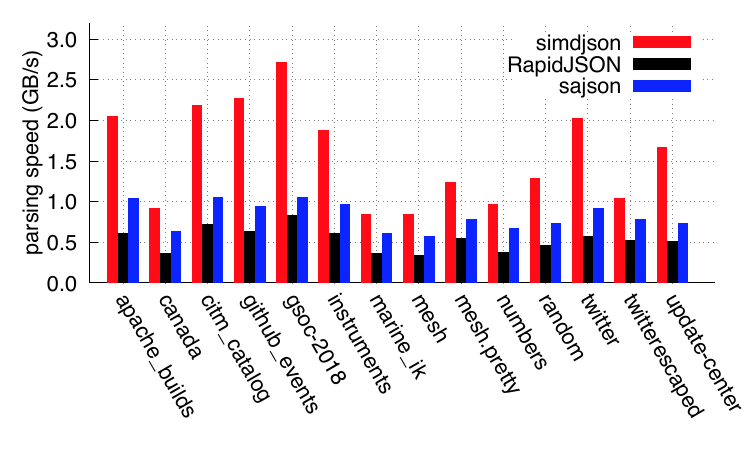}}
\caption{\label{fig:speed}Speed of the three parsers (simdjson, RapidJSON and sajson) while parsing our different files (in GB/s). }
\end{figure*}

We present raw parsing speeds in  Fig.~\ref{fig:speed}.
In only a few instances, sajson can slightly surpass 1\,GB/s and RapidJSON can slightly surpass 0.5\,GB/s. On our Skylake (3.4\,GHz) processor, our parser (simdjson) can achieve and even surpass 2\,GB/s in six instances, and for gsoc-2018, we  reach 3\,GB/s.

The purpose of parsing is to access the data contained in the document. It would not be helpful to quickly parse documents if we could not, later on, access the parsed tree quickly.
In Table~\ref{tab:userid}, we present our results while parsing the twitter document and finding all unique user.id (SELECT DISTINCT “user.id” FROM tweets), a query from Tahara et al.~\cite{Tahara:2014:SSS:2588555.2612183}. We report the throughput in GB/s to parse and scan the resulting tree. Our parser is again twice as fast as the reference parsers.
We fully parse and then fully traverse the result, in two stages, without any optimization. We  achieve 1.8\,GB/s, down from 2.2\,GB/s, when merely parsing on our Skylake processor.
Unsurprisingly, a full parsed-tree traversal takes a fraction of the time required to fully parse the original document---reaching the equivalent of 10\,GB/s in our case. 

\begin{table}\centering
\caption{ \label{tab:userid}Throughput in GB/s to  parse and then  select all distinct \texttt{user.id} values from the parsed tree, using the file twitter. We provide corresponding time for merely parsing and generator a parsed tree, or merely traversing the parsed tree looking for distinct \texttt{user.id} values.}
\subfloat[Skylake]{
\begin{tabular}{lrrr}\toprule
parser & parse + select & parse only & select only \\\midrule
simdjson   & 1.8 & 2.2 & 11.7 \\ 
RapidJSON & 0.7 & 0.8 & 10.8\\
sajson & 0.9 & 1.0 & 15.5\\ \bottomrule
\end{tabular}
}
\hspace{0.1\textwidth}
\subfloat[Cannon~Lake]{
\begin{tabular}{lrrr}\toprule
parser & parse + select & parse only & select only \\\midrule
simdjson   & 1.6 & 2.0 & 9.4\\ 
RapidJSON & 0.6 & 0.7 & 9.7 \\
sajson & 0.8 & 0.9 & 13.6 \\ \bottomrule
\end{tabular}
}
\end{table}


%
%
%
%
%

In Table~\ref{tab:broadcomparison}, we present the
parsing speed in gigabytes per second (GB/s) for several different parsers. In particular, we present results regarding RapidJSON using both the default configuration and the faster version that we use elsewhere (with insitu processing). In several cases, insitu processing is faster (apache\_build, gsoc-2018, twitter, etc.)  up to a factor of two, while in other cases, such as all files made mostly of numbers, the difference is negligible.
Compared with RapidJSON without insitu string processing, our parser (simdjson) can be more than five times faster.
RapidJSON further allows us to disable character encoding validation or to use higher precision number parsing, we do not report the results for these cases. For sajson, we use both the default dynamic-memory allocation and the faster version with static-memory allocation which we use elsewhere. The dynamic-memory allocation leads to a significant performance penalty, but we expect that it makes the parser more conservative in its memory usage. For reference, we also include several other popular C/C++  parsers even though we found them all to be lacking regarding their validation:
the Dropbox parser\footnote{\url{https://github.com/dropbox/json11}}, 
fastjson\footnote{\url{https://github.com/mikeando/fastjson}},
gason\footnote{\url{https://github.com/vivkin/gason}},
ujson4c: a wrapper around the UltraJSON library\footnote{\url{https://github.com/esnme/ujson4c}},
jsmn\footnote{\url{https://github.com/zserge/jsmn}},
 cJSON\footnote{\url{https://github.com/DaveGamble/cJSON}}, and
 jsoncpp\footnote{\url{https://github.com/open-source-parsers/jsoncpp}}, and ``JSON for Modern C++''\footnote{\url{https://nlohmann.github.io/json/}}.
 In all cases, we used the latest available version and we tried to benchmark to get the best speed. Out of these other parsers, the most competitive regarding speed is gason, as it is close to the best performance of sajson.

\begin{table*}\centering
\caption{ \label{tab:broadcomparison}Parsing speed in gigabytes per second (GB/s) for several different parsers (Skylake) }\small
\newcommand*\rot{\rotatebox{90}}
\begin{tabular}{ll|llll|llllllll}
 	&	 \rot{simdjson} 	&	 \rot{RapidJSON} 	&	 \rot{RapidJSON ins.} 	&	 \rot{sajson dyn.} 	&	 \rot{sajson} 	&	 \rot{Dropbox} &	 \rot{fastjson} 	&	 \rot{gason} 	&	 \rot{ultrajson} 	&	 \rot{jsmn}	&	 \rot{cJSON}	&	 \rot{jsoncpp} 	&	 \rot{\pbox{2cm}{JSON for\\ Modern C++}}  \\\midrule
apache_builds	&	2.3	&	0.48	&	0.80	&	0.84	&	1.2	&	0.16	&	0.28	&	0.91	&	0.37	&	0.07	&0.29	&	0.15 & 0.10\\
canada	&	1.1	&	0.43	&	0.43	&	0.60	&	0.76	&	0.07	&	0.22	&	0.95	&	0.48	&	0.01	&	0.07	&	0.04 & 0.05\\
citm_catalog	&	2.5	&	0.86	&	1.0	&	0.85	&	1.2	&	0.24	&	0.39	&	1.1	&	0.65	&	0.19	&0.36	&	0.20 & 0.14\\
github_events	&	2.5	&	0.46	&	0.79	&	1.0	&	1.1	&	0.15	&	0.28	&	0.89	&	0.38	&	0.52	&0.28	&	0.14 & 0.09\\
gsoc-2018	&	3.2	&	0.51	&	1.0	&	0.96	&	1.1	&	0.23	&	0.31	&	1.1	&	0.44	&	0.16	&0.53	&	0.26 & 0.12\\
instruments	&	2.1	&	0.59	&	0.76	&	0.70	&	1.1	&	0.14	&	0.34	&	0.95	&	0.43	&	0.24	&0.25	&	0.13 & 0.10\\
marine_ik	&	0.94	&	0.45	&	0.45	&	0.51	&	0.67	&	0.07	&	0.22	&	0.77	&	0.37	&	0.18	&0.07	&	0.04 & 0.06\\
mesh	&	0.95	&	0.41	&	0.41	&	0.52	&	0.62	&	0.09	&	0.20	&	0.77	&	0.38	&	0.05	&	0.06	&	0.04 & 0.06\\
mesh.pretty	&	1.5	&	0.67	&	0.67	&	0.64	&	0.92	&	0.15	&	0.32	&	1.1	&	0.62	&	0.09	&0.13	&	0.07& 0.11\\
numbers	&	1.1	&	0.45	&	0.45	&	0.55	&	0.77	&	0.09	&	0.24	&	0.88	&	0.48	&	0.65	&	0.07	&	0.04 & 0.06\\
random	&	1.4	&	0.38	&	0.58	&	0.46	&	0.81	&	0.10	&	0.26	&	0.79	&	0.30	&	0.03	&	0.18	&	0.09 & 0.07\\
twitter	&	2.2	&	0.45	&	0.72	&	0.73	&	1.0	&	0.14	&	0.28	&	0.80	&	0.42	&	0.27	&	0.33	&	0.14 & 0.10\\
twitterescaped	&	1.2	&	0.38	&	0.62	&	0.62	&	0.84	&	0.11	&	0.26	&	0.72	&	0.36	&	0.26	&0.27	&	0.11 & 0.08\\
update-center	&	1.9	&	0.38	&	0.64	&	0.50	&	0.86	&	0.11	&	0.25	&	0.71	&	0.31	&	0.06	&0.25	&	0.10 & 0.07\\\bottomrule
\end{tabular}
\end{table*}

\section{Conclusion and Future Work}

Though the application of SIMD instructions for parsing is not novel~\cite{Cameron:2008:HPX:1463788.1463811}, our results suggest that they are underutilized in popular JSON parsers.
We expect that many of our strategies could benefit existing JSON parsers like RapidJSON\@. It may even be possible to integrate the code of our parser (simdjson) directly into existing libraries.



JSON is one of several popular data formats such as Protocol Buffers,  XML, YAML, MessagePack, BSON, CSV, or CBOR\@.  We expect that many of our ideas would apply to other formats.

JSON documents are all text. Yet we frequently need to 
embed binary content inside such documents. The standard
approach involves  using base64 encoding. Base64 data can be decoded quickly using SIMD instructions~\cite{simdbase64}.
Because number parsing from text is expensive, it might be
fruitful to store large arrays of numbers in binary format using base64.


Intel has produced a new family of instruction sets with wider vector registers and more powerful instructions (AVX-512). Our Cannon~Lake processor supports these instructions, including the AVX512-VBMI extension, which is relevant to the byte processing required for this work. 
Future research should assess the benefits of AVX-512 instructions.



Many of our strategies are agnostic to the specific architecture of the processor. Future research should assess our performance  on other processors, such as those of the ARM or POWER families.\footnote{The simdjson library works on 64-bit ARM processors.}

In our current implementation of simdjson, we complete the entire stage~1 before completing stage~2. For large documents, we could interleave stage~1 and stage~2. If needed, stage~1 can be further parallelized with thread-level parallelism, as the bulk of the work done within stage~1 requires only local knowledge of the input.





\begin{acknowledgements}
The vectorized UTF-8 validation was motivated
by a blog post by O.~Goffart. K.~Willets helped design the current vectorized UTF-8 validation. In particular, he provided the algorithm and code to check that sequences of two, three and four non-ASCII bytes match the leading byte.
The authors are grateful to
 W.~Mu\l{}a for sharing related number-parsing code online. The software library has benefited
 from the contributions of  T.~Navennec,  K.~Wolf,  T.~Kennedy, F.~Wessels, G.~Fotopoulos, H.~N.~Gies, E.~Gedda, G.~Floros, D.~Xie, N.~Xiao, E.~Bogatov, J.~Wang, L.~F.~Peres, W.~Bolsterlee, A.~Karandikar, R.~Urban, T.~Dyson, I.~Dotsenko, A.~Milovidov, C.~Liu, S.~Gleason, J.~Keiser, Z.~Bjornson, V.~Baranov, I.~A.~Daza Dillon and others.

The work is supported in part by the 
Natural Sciences and Engineering Research Council of Canada 
under grant RGPIN-2017-03910. 
\end{acknowledgements}
\appendix
\section{Effect of Minification}

To ease readability, JSON documents may contain a variable number of white-space characters between atoms, within objects and arrays. Intuitively, these superfluous characters should reduce parsing speed. To verify this intuition, we \emph{minified}  the documents prior to parsing. We find that all three parsers (simdjson, RapidJSON, sajson) use fewer CPU cycles to parse minified documents, see Table~\ref{tab:minification}. However, the benefits  in processing speed is often less than the benefits in  storage. For example, the minified apache_builds file is 74\% of the original,  yet the processing time is only reduced to between 82\% and 90\% of the original---depending on the parser. The sajson parser often benefits more from minification.  Thus while simdjson is more than 2.1~times faster than sajson on the original twitter document, it is only 1.9~times faster after minification.

\begin{table*}
\caption{\label{tab:minification}Millions of cycles required to parse and validate selected  documents (Skylake), before and after minification.
}
\centering
\begin{minipage}{\textwidth}
\centering
\begin{tabular}{lccccc}\toprule
 file &  ratio (\%) & parser & original & minified  & $\frac{\mathrm{original}}{\mathrm{minified}}/$ (\%)\\
\midrule
apache_builds &       74 &                   &         &   & \\
              &          & simdjson.         &   0.19  & 0.17& 86\\
& & RapidJSON                                 &   0.55  & 0.49& 90\\
& & sajson                                 &   0.34  & 0.28& 82\\
 citm_catalog &       29  &&  &  & \\
 & &simdjson                               &   2.34  & 1.64& 70\\
 & &RapidJSON                                  &   6.30  & 3.49& 55\\
 & &sajson                                 &   4.79  & 2.40& 50\\
github_events  &       82 &&    & & \\
 & &simdjson                                &   0.09  & 0.08& 90\\
 & &RapidJSON                                  &   0.29 &  0.25& 86\\
 & &sajson                                 &   0.20  & 0.16& 81\\
gsoc-2018 &       92 & &  &  & \\
 & &simdjson                                &   3.58  & 3.39& 95\\
 & &RapidJSON                                  &  11.2 & 10.6& 95\\
 & &sajson                                  &   8.76 &  8.19& 95\\
instruments &       49 & &    & & \\
 & &simdjson                                &   0.36 &  0.29 & 80\\
 & &RapidJSON                                  &   0.98 &  0.75& 77\\
 & &sajson                                  &   0.70  & 0.47& 68\\
marine_ik &       61 & &  &  & \\
 & &simdjson                                &  10.6&  10.2& 96\\
 & &RapidJSON                              &  22.3&  20.2& 91\\
 & &sajson                                &  15.1 & 12.9& 85\\
mesh  &       90  & &  &   & \\
 & &simdjson                              &   2.56  & 2.54& 99\\
 & &RapidJSON                                 &   6.06 &  6.00& 99\\
 & &sajson                                 &   3.88 &  3.70& 96\\
mesh.pretty &       48 &   & & & \\
 & &simdjson                                &   3.64 &  3.19& 88\\
 & &RapidJSON                                 &   8.50 &  6.56& 77\\
 & &sajson                                  &   5.61  & 4.08& 73\\
random &       90  & &  & & \\
 & &simdjson                               &   1.22 &  1.19& 97\\
 & &RapidJSON                              &   2.96 &  2.90& 98\\
 & &sajson                                 &   2.08 &  1.96& 94\\
twitter   &       74 &  & & & \\
 & &simdjson                               &   0.99 &  0.88& 90\\
 & &RapidJSON                              &   3.08 &  2.40& 78\\
 & &sajson                                 &   2.11  & 1.63& 77\\
\bottomrule
\end{tabular}
\end{minipage}
\end{table*}

\section{Effect of Specific Optimizations}

In a complex task like JSON parsing, no single optimization is likely to make a large difference. Nonetheless, it may be useful  to quantify the effect of some optimizations.
\begin{itemize}
\item We can compute the string masks (indicating the span of the strings) using a single carry-less multiplication between a word containing the location of the quote characters (as 1-bits) and a word containing only ones (see \S~\ref{sec:escapedquotes}). Alternatively, we can achieve the same result with a series of shifts and XOR: 
\lstset{escapechar=@,style=customc}
\begin{lstlisting}
uint64_t mask = quote xor (quote << 1);
mask = mask xor (mask << 2);
mask = mask xor (mask << 4);
mask = mask xor (mask << 8);
mask = mask xor (mask << 16);
mask = mask xor (mask << 32);
\end{lstlisting}
\item Instead of extracting the set bits using our optimized algorithm (see Fig.~\ref{fig:extraction} in \S~\ref{sec:indexextract}), we can use a naive approach:
\lstset{escapechar=@,style=customc}
\begin{lstlisting}
while(s != 0) {
  *b++ = idx + trailingzeroes(s);
  s = s & (s - 1);
}
\end{lstlisting}
\item Instead of using vectorized classification (\S~\ref{sec:vectorizedclassification}), we can use a more 
naive approach where we detect the locations of structural characters by
doing one vectorized comparison per structural character, and doing a bitwise OR\@. Similarly, we can detect spaces by doing one comparison per allowable white-space character and doing a bitwise OR.
\end{itemize}
We present the number of cycles per input byte in Table~\ref{tab:optimization} with the three optimizations disabled one by one.
The standard error of our measure is about 0.02~cycles per byte so small differences ($<0.05$~cycles) may not be statistically significant. The fast index extraction reduces the cost of the stage~1 by over 10\% in several instances (e.g., twitter, update-center). The carry-less multiplication appear has gains of over 5\% in some instance, the gains reach nearly 20\% in one instance (mesh). Vectorized classification is similarly helpful.

\begin{table*}
\caption{\label{tab:optimization}Performance in cycles per bytes of the simdjson parser during stage~1 over several files. The first numerical column presents the original performance, with all optimizations. The second column presents the results without the carry-less multiplication. The third column presents the results with a naive bit extraction. The fourth column presents the results with naive classification---without vectorized classification.
}
\centering
\begin{minipage}{\textwidth}
\centering
\begin{tabular}{lc|ccc}\toprule
 file &  original & without carry-less & naive bit extraction & naive classication\\
\midrule
canada        &   0.90 & 0.94 & 0.96 & 0.96\\
 citm_catalog &    0.75& 0.78 & 0.80 & 0.81\\
github_events  &   0.74 & 0.77 &0.74 & 0.83\\
gsoc-2018 &       0.65 & 0.67 & 0.69 & 0.68\\
instruments &     0.81 & 0.84 &1.01 & 0.88\\
marine_ik &       1.02 &1.09& 1.25 & 1.09\\
mesh  &            0.90& 1.07 & 1.07 & 0.99\\
mesh.pretty &     0.77& 0.82 & 0.82 & 0.85\\
numbers         & 0.83& 0.86 &0.88 & 0.89\\
random &        1.22& 1.26& 1.57 & 1.31\\
twitter   &  0.90 &  0.92 &1.14 & 1.00\\
twitterescaped &0.79& 0.86& 0.96 & 0.85\\
update-center &0.85 &0.87 &1.09 & 0.91\\
\bottomrule
\end{tabular}
\end{minipage}
\end{table*}

\section{Large Files}

All three fast parsers (simdjson, RapidJSON, sajson) mostly read and write sequentially in memory. Furthermore, the memory bandwidth of our systems are far higher than our parsing speeds. Thus we can expect them to perform well even when all of the data does not fit in cache. To verify that we can still process JSON data at high speed when the input data exceeds the cache, we created a large file (refsnp-unsupported35000) made of the first
35,000~entries from a file describing human single nucleotide variations (refsnp-unsupported.json). Table~\ref{tab:largefile} shows that we far exceed 1\,GB per second with simdjson in this instance  although the file does not fit in CPU cache.

\begin{table}\centering
\caption{ \label{tab:largefile}Throughput in GB/s when parsing a large (84\,MB) file: refsnp-unsupported35000}
\begin{tabular}{lrr}\toprule
parser & Skylake & Cannon Lake \\\midrule
simdjson   & 1.4 & 1.3\\ 
RapidJSON & 0.56 & 0.44\\
sajson & 0.93 & 0.84\\ \bottomrule
\end{tabular}
\end{table}

\bibliographystyle{spbasic}
\bibliography{simdjson-bibliography}

\end{document}